\documentclass[conference]{IEEEtran}
\IEEEoverridecommandlockouts
\usepackage[table]{xcolor}
\usepackage{cite}
\usepackage{amsmath,amssymb,amsfonts}
\usepackage{algorithmic}
\usepackage{graphicx}
\usepackage{textcomp}
\usepackage{xcolor}

\usepackage{multirow}

\usepackage{bm}
\usepackage{booktabs}
\usepackage{threeparttable}

\usepackage{marvosym}

\def\BibTeX{{\rm B\kern-.05em{\sc i\kern-.025em b}\kern-.08em
    T\kern-.1667em\lower.7ex\hbox{E}\kern-.125emX}}
\begin{document}

\title{CA-Diff: Collaborative Anatomy Diffusion for Brain Tissue Segmentation
\thanks{\Letter Corresponding authors: skyesong@hust.edu.cn, weiyangcs@hust.edu.cn}}

\author{\IEEEauthorblockN{Qilong Xing, Zikai Song\textsuperscript{\Letter}, Yuteng Ye, Yuke Chen, Youjia Zhang, Na Feng, Junqing Yu, Wei Yang\textsuperscript{\Letter}}
\IEEEauthorblockA{\textit{School of Computer Science and Technology}, 
\textit{Huazhong University of Science and Technology}, Wuhan, China \\
\{qlxing, skyesong, yuteng\_ye, yuke\_chen, youjiazhang, fengna, yjqing, weiyangcs\}@hust.edu.cn}}

\maketitle
\begin{abstract}
Segmentation of brain structures from MRI is crucial for evaluating brain morphology, yet existing CNN and transformer-based methods struggle to delineate complex structures accurately. While current diffusion models have shown promise in image segmentation, they are inadequate when applied directly to brain MRI due to neglecting anatomical information. To address this, we propose Collaborative Anatomy Diffusion (CA-Diff), a framework integrating spatial anatomical features to enhance segmentation accuracy of the diffusion model. Specifically, we introduce distance field as an auxiliary anatomical condition to provide global spatial context, alongside a collaborative diffusion process to model its joint distribution with anatomical structures, enabling effective utilization of anatomical features for segmentation. Furthermore, we introduce a consistency loss to refine relationships between the distance field and anatomical structures and design a time adapted channel attention module to enhance the U-Net feature fusion procedure. Extensive experiments show that CA-Diff outperforms state-of-the-art (SOTA) methods.
\end{abstract}
    

\section{Introduction}
\label{sec:intro}

Segmentation of brain structures from Magnetic Resonance Imaging (MRI) plays an indispensable role across a spectrum of clinical applications, including tracking of the development and disease diagnosis of the brain \cite{akkus2017deep}.
The trend towards automation of brain structures segmentation has been propelled by strides in deep learning technologies,
especially Convolutional Neural Networks (CNNs)~\cite{nnunet,zhou2018unet++} and Vision Transformers (ViTs)~\cite{swinunetr,unetr,song2022transformer,song2023compact}.
However, these methods primarily approach segmentation from a discriminative perspective, encountering difficulties in accurately segmenting complex structures, with limited investigation into generative approaches.

Recently, diffusion and score-based methods have demonstrated significant performance across a variety of generation tasks \cite{luo2024diffusiontrack,zhou2024video,qiu2023diffbfr}, showcasing their ability to iteratively refine predictions. 
The remarkable efficacy of these methods has spurred growing interest in extending the application of diffusion models to generate segmentation labels for medical images \cite{diffunet,cDAL,corrdiff}.
From the diffusion method perspective, segmentation can be
regarded as a label map generation procedure with the image itself as conditioning factor. 
Nevertheless, these methods often experience a decline in performance as they overlook the knowledge embedded in anatomical features.

\begin{figure}[t]
  \centering
    \includegraphics[width=0.8\linewidth]{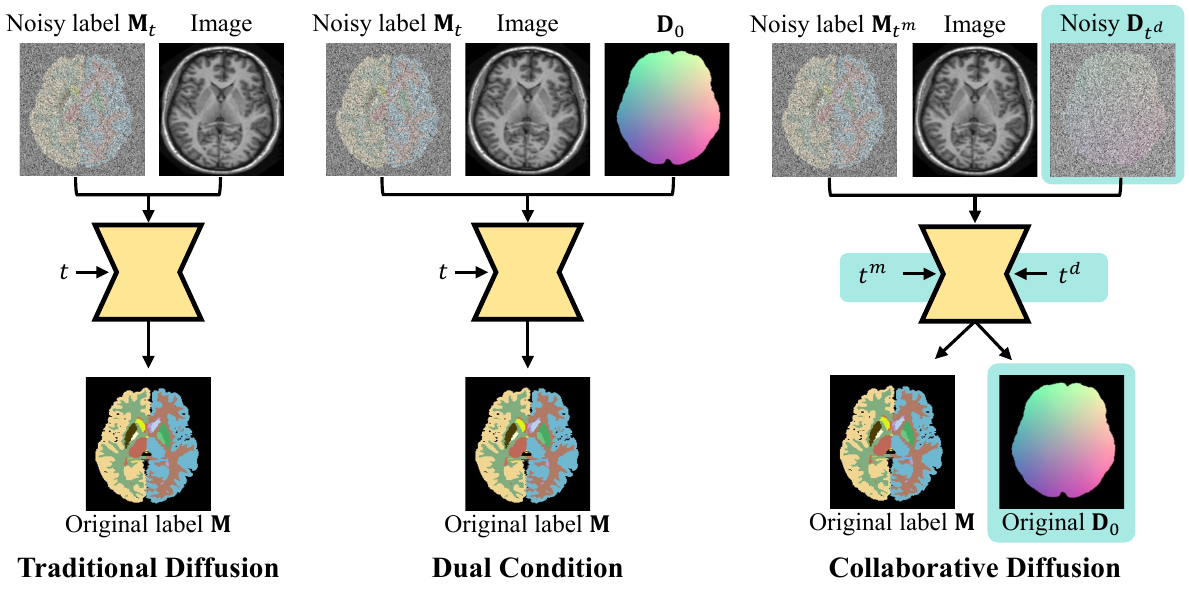}
    \caption{
    While \textit{Traditional Diffusion} method are adapted for MRI label generation solely based on image, the \textit{Dual Condition} is able to use the anatomical data as additional condition, which is the distance field in our case. We introduce the CA-Diff framework to utilize the \textit{Collaborative Diffusion} approach to collaboratively learns to denoise both the semantic label map and the distance field to use the anatomical data more effectively. Here we denote the original input distance field as $\mathbf{D}_0$ and the noisy version at time $t^d$ as $\mathbf{D}_{t^d}$. The details about the distance field are described in the Method section. 
    }
    \label{fig:compare}
  \hfill
\end{figure}

Anatomical features commonly used in segmentation tasks include shape information and appearance information~\cite{Liu_Wolterink_Brune_Veldhuis_2021}. 
For brain MRI, probabilistic atlases \cite{mazziotta1995probabilistic} provide valuable anatomical information by containing the probability of occurrence of brain structures at different locations. However, probabilistic atlases, which primarily contain local information, are insufficient for assisting segmentation procedures from a global perspective. 
Additionally, the method of integrating anatomical features with the diffusion model is crucial. 
To incorporate the anatomical feature into the diffusion model, a straightforward approach is to use them as an additional condition by simply combining them with the image, 
which can be termed as the \textit{Dual Condition} method (Fig. \ref{fig:compare}). 
However, this method fails to fully utilize the anatomical feature to enhance segmentation performance.

To this end, we introduce the Collaborative Anatomy Diffusion (CA-Diff) framework, which generates labels using a diffusion model guided by both the image and the extracted anatomical features for brain structures from MRI scans.
To enhance the model's understanding of the entire brain's spatial features, we propose utilizing global spatial information through a distance field to assist in the segmentation process.
To effectively utilize the distance field as the auxiliary condition, we propose a collaborative diffusion method to better connect the distance field and anatomical structures, rather than simply using distance field as an additional condition (Fig. \ref{fig:compare}).
Specifically, CA-Diff perturbs the distance field and label with Gaussian noise and denoises them collaboratively, inspired by UniDiffuser \cite{unidiffuser} on cross modality learning. 
Our approach fully explores the association between the spatial locations and anatomies, enhancing model performance compared to the \textit{Dual Condition} method (Fig.~\ref{fig:compare}). 
Furthermore, we introduce a consistency loss to strengthen the association between the distance field and anatomical structures and propose a Time-Adapted Channel Attention (TACA) mechanism to improve feature fusion in the diffusion U-Net's skip connections.

\section{Method}
We propose reinterpreting the segmentation problem as a label generation task and introduce the CA-Diff framework to enable this process. The overall workflow of our approach is illustrated in Fig.~\ref{fig:overall_arch}.

\subsection{Segmentation via Diffusion Model.}
Diffusion models transform samples from Gaussian noise into samples that align with an empirical data distribution. The training of diffusion models involves two stages to learn this transformation: a forward process, where the noise is added to the samples, and a reverse denoising procedure carried out by a neural network $\phi_{\theta}$.
For image segmentation task, diffusion models can be adapted into a conditional generative process for generating the label $\mathbf{M}_0$. Specifically, in the forward process, noise is initially added to the original label to obtain the perturbed version at time $t$:
\begin{equation}
  \mathbf{M}_t=\sqrt{\alpha_{t}} \mathbf{M}_{0}+\sqrt{1-\alpha_{t}} \epsilon
  \label{eq:forward_noise}
\end{equation}
where $\epsilon$ is drawn from starndard normal density, $\alpha_{t}$ are defined by specific noise schedule and $t\in\{0,1,...,T\}$.
As for the reverse denoising procedure, the corresponding image $\mathbf{I}$ is commonly used as the condition
and the final label $\mathbf{M}_0$ is generated by $\phi_{\theta}$ from a random noise $\mathbf{M}_T$ in an iterative markovian manner:
\begin{equation}
  p_{\theta}(\mathbf{M}_{0:T}|\mathbf{I})=p(\mathbf{M}_T)\prod_{t=1}^{T}p_{\theta}(\mathbf{M}_{t-1}|\mathbf{M}_{t},\mathbf{I}).
  \label{eq:denoise}
\end{equation}
During training, $\phi_{\theta}$ learns to predict the label $\mathbf{M}_0$ from $\mathbf{M}_t$ and the segmentation loss has been widely used as the objective function.

\subsection{Anatomical Distance Field}
Contrasting with natural images, medical images typically exhibit a more consistent spatial structure, which allows for the identification of anatomical structures based on their positions. 
Building on this observation, we propose using the distance field to encode the spatial relations of brain structures and act as additional anatomical guidance in the label generation procedure.
Specifically, the distance field is initially constructed in the atlas image space. We select the origin point at the anterior commissure and equally divide the brain into left and right hemispheres to establish the coordinate system. The distance field contains the 3D coordinates for each voxel of the brain area, providing a granular representation of brain spatial information. To obtain the distance field for each sample, we employ registration method to map the distance field from atlas space to sample space. This registration method ensures that similar anatomical structures in different sample images are consistently located in the atlas space, thereby facilitating the association of spatial locations with anatomies. Consequently, the distance field can provide global spatial information to guide the label generation procedure of the diffusion model while alleviating the global information loss caused by the patch-based training method. 

\begin{figure*}[t]
  \centering
    \includegraphics[width=0.65\linewidth]{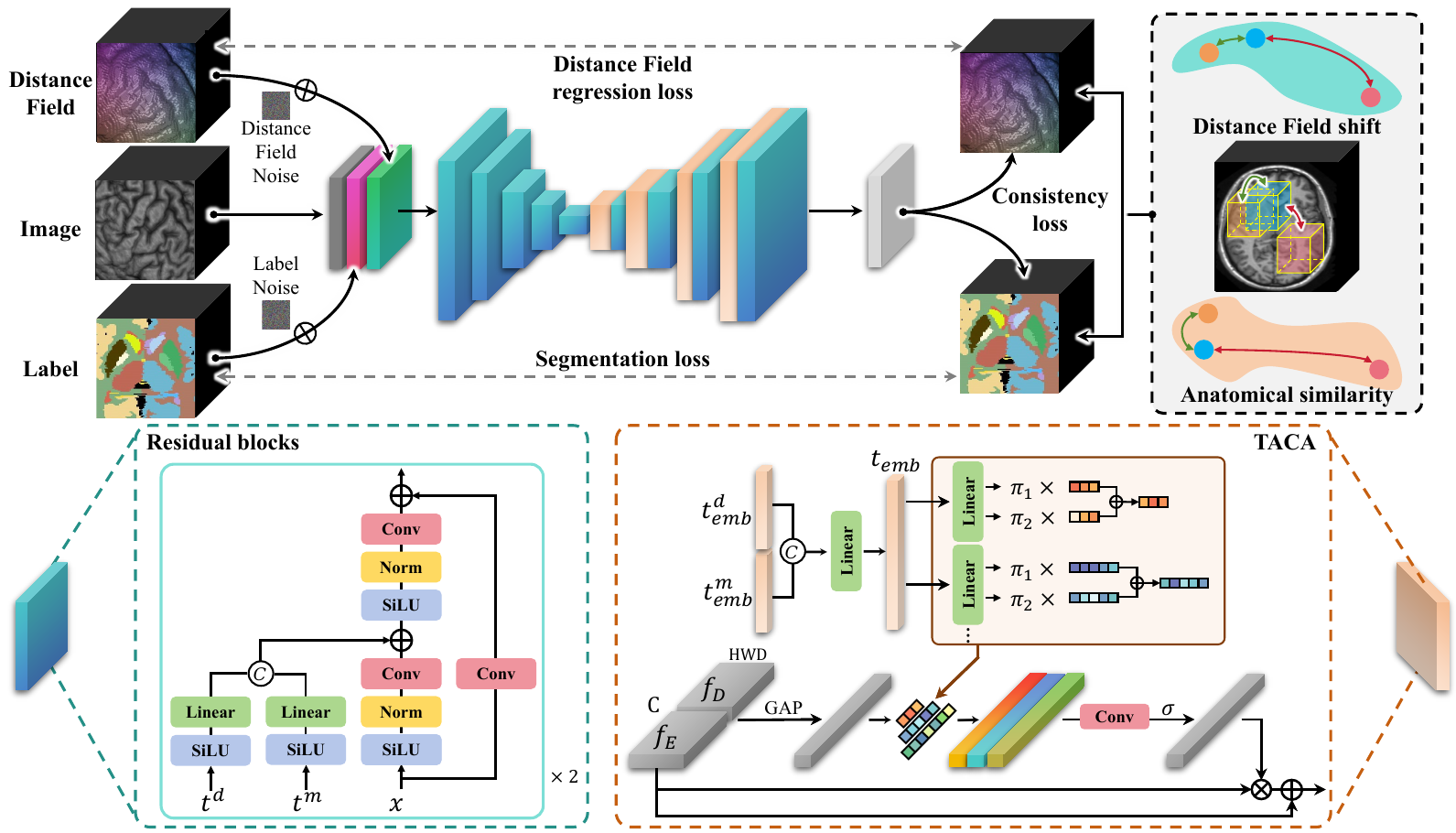}
    \caption{The pipeline of our CA-Diff.
The distance field and label are perturbed using Gaussian noise, and $\phi_{\theta}$ is used to generate the original distance field and label. $\phi_{\theta}$ is constructed using residual blocks, each containing separate linear layers to obtain time embeddings for the distance field and label, respectively. Supervision of $\phi_{\theta}$ is provided by the segmentation loss, the distance field regression loss, and a novel consistency loss, which is designed to refine the relationship between the distance field and label. To better fuse features in skip connections, a time-adapted channel attention module is proposed, where dynamic convolution is employed to determine the attention weights according to the current diffusion step.
    }
    \label{fig:overall_arch}
  \hfill
\end{figure*}

\subsection{Collaborative Anatomy Diffusion}
Formally, we use the extracted sample distance field $\mathbf{D}_0$ and the input image $\mathbf{I}$ to guide the diffusion model in generating the final segmentation label $\mathbf{M}_0$. 
In contrast to the \textit{Dual Condition} method, which fails to effectively leverage the anatomical information embedded within its framework by focusing solely on learning $q(\mathbf{M}_0|\mathbf{D}_0,\mathbf{I})$, we propose an enhanced approach that additionally learns the joint distribution $q(\mathbf{D}_0,\mathbf{M}_0|\mathbf{I})$.
In the context of diffusion models, we are able to estimate a conditional expectation over the original inputs to learn the distribution for our segmentation task. 
Specifically, the distribution $q( \mathbf{M}_0 | \mathbf{D}_0, \mathbf{I})$ can be modeled by estimating the expectation over the original input label $\mathbf{M}_0$, i.e., $\mathbb{E}[\mathbf{M}_0|\mathbf{D}_0,\mathbf{M}_t,\mathbf{I}]$, while the distribution $q(\mathbf{D}_0,\mathbf{M}_0|\mathbf{I})$ can be represented by estimating $\mathbb{E}[\mathbf{D}_0,\mathbf{M}_0|\mathbf{D}_t,\mathbf{M}_t,\mathbf{I}]$.

To learn multiple distributions using a single model, we propose the Collaborative Anatomy Diffusion (CA-Diff) framework, which is designed to consider multiple potential distributions through learning the general form distribution $\mathbb{E}[\mathbf{D}_0,\mathbf{M}_0|\mathbf{D}_{t^{d}},\mathbf{M}_{t^{m}},\mathbf{I}]$, where $t^{d}$ and $t^{m}$ are two distinct timesteps, and $\mathbf{D}_{t^{d}}$ and $\mathbf{M}_{t^{m}}$ represent the perturbed versions of the distance field and label, respectively. 
The choice of timesteps in $\mathbb{E}[\mathbf{D}_0,\mathbf{M}_0|\mathbf{D}_{t^{d}},\mathbf{M}_{t^{m}},\mathbf{I}]$ leads to different distributions. By setting $t^{d}=t^{m}=T$, we have $\mathbb{E}[\mathbf{M}_0|\mathbf{D}_{T},\mathbf{M}_{T},\mathbf{I}]\approx \mathbb{E}[\mathbf{M}_0|\mathbf{I}]$, which is the direct generation of the label solely relying on the target image. By setting $t^{d}=0$ and $t^{m}=T$, we derive $\mathbb{E}[\mathbf{M}_0|\mathbf{D}_0,\mathbf{M}_{T},\mathbf{I}]\approx \mathbb{E}[\mathbf{M}_0|\mathbf{D}_0,\mathbf{I}]$, where the distance field is used as an additional condition to guide the label generation procedure.
To learn the joint distribution, we can set $t^{d}=t^{m}=t$ to learn $\mathbb{E}[\mathbf{D}_0,\mathbf{M}_0|\mathbf{D}_t,\mathbf{M}_t,\mathbf{I}]$, which helps the model learn the spatial features of the brain and construct the relationship between spatial locations and anatomical regions. Moreover, learning $\mathbb{E}[\mathbf{D}_0,\mathbf{M}_0|\mathbf{D}_t,\mathbf{M}_t,\mathbf{I}]$ can be extended by setting $t^{d}>0$ and $t^{m}>0$ with different values, which can serve as an additional augmentation method for better relating $\mathbf{D}_0$ to $\mathbf{M}_0$. 

To learn $\mathbb{E}[\mathbf{D}_0,\mathbf{M}_0|\mathbf{D}_{t^{d}},\mathbf{M}_{t^{m}},\mathbf{I}]$, CA-Diff employs a collaborative diffusion U-Net model $\phi_{\theta}$ to accept the perturbed versions of $\mathbf{D}_0$ and $\mathbf{M}_0$ and predicts the original inputs collaboratively. The architecture is illustrated in Fig. \ref{fig:overall_arch}. $\mathbf{D}_0$ and $\mathbf{M}_0$ are perturbed and input to $\phi_{\theta}$ via concatenation with the target image. 
The core module of $\phi_{\theta}$ consists of residual blocks that utilize separate linear layers to independently learn the time embeddings corresponding to the distance field and label. These embeddings are then concatenated along the channel dimension and added to the feature map.
The final layer of $\phi_{\theta}$ has two branches, one for the regression of $\mathbf{D}_0$ and the other for the prediction of $\mathbf{M}_0$:
\begin{equation}
\hat{\mathbf{D}}_0, \hat{\mathbf{M}}_0= \phi_{\theta} (\mathbf{D}_{t^d}, \mathbf{M}_{t^{m}}, \mathbf{I}, t^{d}, t^m)
\end{equation}
Here, the timesteps $t^{d}$ and $t^m$ are uniformly and independently sampled from $\{0,1,...,T\}$ to obtain the noisy inputs $\mathbf{D}_{t^d}$ and $\mathbf{M}_{t^{m}}$.
To generate $\mathbf{D}_0$ and $\mathbf{M}_0$ simultaneously, we propose to minimize the following loss for $\phi_{\theta}$:
\begin{equation}
\begin{aligned}
\mathcal{L}_{\mathrm{diff}}&= \mathbf{MSE}(\hat{\mathbf{D}}_0,\mathbf{D}_0) \\
&+\mathbf{Dice}(\hat{\mathbf{M}}_0,\mathbf{M}_0)+\mathbf{CE}(\hat{\mathbf{M}}_0,\mathbf{M}_0)
  \label{eq:diff_loss}
\end{aligned}
\end{equation}
where the loss $\mathcal{L}_{\mathrm{diff}}$ comprises the mean squared error (MSE) loss for distance field regression, the Dice loss (Dice) and the cross-entropy loss (CE) for label generation. 

CA-Diff captures the complex relationships between space and anatomies and enhances model understanding of spatial feature, thus fully utilizes the distance field to improve segmentation accuracy. 
During inference, we focus solely on generating the segmentation label, using the clean distance field as the condition with $t^{d}=0$.

\begin{table*}[t]
\caption{Performance comparison on three datasets. All methods are rerun using \\three-fold cross-validation without post-processing to ensure a fair comparison.}\label{result_compare}
\centering
\renewcommand\arraystretch{1.0}
\scalebox{1}{
\begin{tabular}{p{100pt}<{\centering} p{40pt}<{\centering} p{40pt}<{\centering} p{40pt}<{\centering} p{40pt}<{\centering} p{40pt}<{\centering} p{40pt}<{\centering} }
\toprule[0.7pt]
\multirow{3}{*}{\textbf{Method}}                      & \multicolumn{6}{c}{\textbf{Class Average Value}}                                                                                                                                           \\ \cline{2-7} 
                                             & \multicolumn{2}{c}{\textbf{MALC}}                                        & \multicolumn{2}{c}{\textbf{SchizBull}}                                   & \multicolumn{2}{c}{\textbf{Hammers}}                 \\ \cline{2-7} 
                                             & Dice$\uparrow$                 & \multicolumn{1}{c}{NSD$\uparrow$}                  & Dice$\uparrow$                 & \multicolumn{1}{c}{NSD$\uparrow$}                  & Dice$\uparrow$                 & NSD$\uparrow$                  \\ \hline
Multi-Atlas                                  & 86.51(0.71)          & \multicolumn{1}{c}{88.42(2.13)}          & 80.69(0.68)          & \multicolumn{1}{c}{88.42(1.65)}          &71.31(0.67)                      &65.87(1.47)                      \\
nnU-Net 2D\cite{nnunet}     & 87.36(0.62)          & \multicolumn{1}{c}{89.63(1.89)}          & 87.27(0.63)          & \multicolumn{1}{c}{88.06(1.97)}          & 77.22(0.54)                     &70.16(1.32)                      \\
UNet++\cite{zhou2018unet++} & 87.51(0.59)          & \multicolumn{1}{c}{89.91(1.95)}          & 87.29(0.67)          & \multicolumn{1}{c}{87.09(1.74)}          & 78.24(0.75)                     &72.18(1.45)                      \\
nnU-Net 3D\cite{nnunet}     & 88.62(0.36)          & \multicolumn{1}{c}{92.65(0.36)}          & 87.93(0.44)          & \multicolumn{1}{c}{89.01(0.57)}          & 81.29(0.53)          & 76.72(1.19)          \\
V-Net\cite{vnet}            & 88.67(0.40)          & \multicolumn{1}{c}{92.79(0.32)}          & 88.04(0.62)          & \multicolumn{1}{c}{89.20(1.18)}          & 81.18(0.68)          & 76.53(1.02)          \\
UNETR\cite{unetr}           & 88.84(0.52)          & \multicolumn{1}{c}{92.12(1.67)}          & 87.57(0.60)          & \multicolumn{1}{c}{88.16(1.42)}          & 81.08(0.94)          & 76.95(1.27)          \\
Swin UNETR\cite{swinunetr}  & 89.39(0.39)          & \multicolumn{1}{c}{92.94(0.37)}          & 88.22(0.41)          & \multicolumn{1}{c}{89.31(1.01)}          & 81.27(0.71)          & 77.62(1.04)          \\
Diff-UNet\cite{diffunet}    & 88.50(0.51)          & \multicolumn{1}{c}{91.83(1.81)}          & 87.85(0.67)          & \multicolumn{1}{c}{88.83(1.74)}          & 81.24(0.82)          & 76.82(0.42)          \\
ACEnet\cite{acenet}         & 89.10(0.46)          & \multicolumn{1}{c}{92.50(0.60)}          & 88.10(0.57)          & \multicolumn{1}{c}{88.97(0.36)}          & 81.35(0.41)          & 78.21(0.56)          \\
CAN\cite{can}               & 89.59(0.44)          & \multicolumn{1}{c}{93.20(0.53)}          & 88.38(0.55)          & \multicolumn{1}{c}{89.23(0.98)}          & 81.42(0.77)          & 77.73(0.45)          \\
\rowcolor{gray!20}
\textbf{CA-Diff}                                & \textbf{90.63(0.41)} & \multicolumn{1}{c}{\textbf{94.32(0.43)}} & \textbf{89.29(0.36)} & \multicolumn{1}{c}{\textbf{89.85(0.80)}} & \textbf{82.54(0.46)} & \textbf{78.52(0.67)} \\ \bottomrule[0.7pt]
\end{tabular}}
\end{table*}

\subsection{Spatial-Anatomical Consistence}
When extracting patches from brain MRIs, patches exhibiting similar anatomical structures are often spatially proximate in the atlas space. Leveraging this insight, we propose a consistency loss to align spatial distance with anatomical similarity, thereby enhancing the relationship between spatial locations and anatomical structures.

Specifically, given two image patches $\mathbf{I}_i$ and $\mathbf{I}_j$, which may be obtained from different MRIs, $\phi_{\theta}$ is used to predict the corresponding $\hat{\mathbf{D}}_0$ and label $\hat{\mathbf{M}}_0$. We compute the similarity based on $\hat{\mathbf{D}}_0$ and $\hat{\mathbf{M}}_0$, denoted as $\mathrm{Sim}_{\mathrm{D}}(i,j)$ and $\mathrm{Sim}_{\mathrm{M}}(i,j)$ for these two samples, and employ the consistency loss to enforce that $\mathrm{Sim}_{\mathrm{D}}(i,j)$ is aligned to $\mathrm{Sim}_{\mathrm{M}}(i,j)$.
To obtain $\mathrm{Sim}_{\mathrm{D}}(i,j)$, 
we use the Euclidean distance to represent the relative shift distance in the atlas space of these two image patches. 
For $\mathrm{Sim}_{\mathrm{M}}(i,j)$, we obtain the predicted logits $\hat{P}$ of the two image patches from the last layer of $\phi_{\theta}$ and apply the cosine similarity function to calculate the anatomical similarity. Finally, the binary cross-entropy loss is employed as the consistency loss $\mathcal{L}_{\mathrm{sac}}(i,j)$:

\begin{equation}
   \mathrm{Sim}_{\mathrm{D}}(i,j)= 1-\dfrac{1}{N} \sum_{n=1}^N \sqrt{\sum_{c \in [0,1,2]} (\hat{\mathbf{D}}_{0,i}^{n,c} - \hat{\mathbf{D}}_{0,j}^{n,c})^2 / 3}
\end{equation}

\begin{equation}
    \mathrm{Sim}_\mathrm{M}(i,j)=\dfrac{1}{N} \sum_{n=1}^N \dfrac{\hat{P}_{i}^{n} \cdot \hat{P}_{j}^{n}}{\Vert \hat{P}_{i}^{n} \Vert \Vert \hat{P}_{j}^{n} \Vert}
\end{equation}

\begin{equation}
\begin{aligned}
\mathcal{L}_{\mathrm{sac}}(i,j) &= -\mathrm{Sim}_{\mathrm{M}}(i,j) \cdot \log (\mathrm{Sim}_{\mathrm{D}}(i,j)) \\
&- (1-\mathrm{Sim}_{\mathrm{M}}(i,j)) \cdot \log (1-\mathrm{Sim}_{\mathrm{D}}(i,j))
\end{aligned}
\end{equation}
where $N$ corresponds to the number of voxels in the cropped image patch, $c$ denotes the channel dimension (which is 3 for 3D images).
While the collaborative diffusion method connects the coordinate values of the distance field to anatomical labels, the consistency loss refines the spatial-anatomical relationship through fine-grained similarity calculations.
Note that values of the distance field are rescaled to between 0 and 1, so the distance shift for each dimension also ranges from 0 to 1.
Thus the overall loss can be formulated as:
\begin{equation}
\mathcal{L}=\mathcal{L}_{\mathrm{diff}}+\mathcal{L}_{\mathrm{sac}}
\end{equation}

\subsection{Time Adapted Channel Attention}
To further improve the segmentation performance of CA-Diff, we introduce a Time Adapted Channel Attention (TACA) module in the skip connections to reweight different features according to the current timestep. This module is based on the premise that the contribution of each channel within a feature map varies at each timestep of the diffusion model \cite{LE_Diff}.

Given the feature $f_E \in \mathbb{R}^{c\times N}$ from the encoder and the upsampled feature $f_D \in \mathbb{R}^{c\times N}$ from the decoder (where $c$ and $N$ refer to channel and number of voxels), we first concatenate them along channel dimension and apply the global average pooling along the spatial dimension in the skip connection:
\begin{equation}
f_{cat}=\mathrm{Cat}([f_E, f_D]), \
f_{pool}=\mathrm{GAP}(f_{cat})
\end{equation}
Following ECA-Net~\cite{eca_net}, we use 1D convolution on the pooled feature map to compute channel attention weights, but with parameters dynamically adjusted based on the current timestep.
Specifically, the time embedding $t_{emb}^{d}$ and $t_{emb}^{m}$ are concatenated along the channel dimension and mapped by a linear layer to get $t_{emb}$:
\begin{equation}
t_{emb}=\mathrm{Linear}( \mathrm{Cat} ([t_{emb}^{d},t_{emb}^{m}]))
\end{equation}
To adaptively modify the convolution parameters, we employ dynamic convolution \cite{dynamic_conv}. This technique allows the attention weights of different convolution kernels to be adjusted based on the current time embedding $t_{emb}$:

\begin{equation}
\begin{split}
&\hat{\pi}=\mathrm{Softmax}( \mathrm{Linear}(t_{emb})) \in \mathbb{R}^K, \\
&\hat{W}=\sum_{k=1}^K \hat{\pi}_k \times W_{k}, \
\hat{b} = \sum_{k=1}^K \hat{\pi}_k \times b_{k} \\
&f = \mathrm{Conv1D}(f_{pool}, \hat{W}, \hat{b})
\end{split}
\end{equation}
where $\hat{\pi}_k$ denotes the attention weight determined by the current timestep for each kernel, while $W_k$ and $b_k$ represent the weight and bias of each learnable kernel used in the dynamic convolution, respectively. The reweighted parameters, $\hat{W}$ and $\hat{b}$, are employed in the 1D convolution layer to incorporate the temporal effect and generate the output feature $f \in \mathbb{R}^{c}$.

To enhance the flexibility of the convolutional operations, we adopt a multi-scale approach by employing multiple groups of dynamic convolutions, each with a different kernel size. The outputs of these dynamic convolution operations are concatenated along the channel dimension and fused by a convolution operation. This output is then used to compute a mask $f_{mask}  \in \mathbb{R}^{c}$ via the sigmoid function, which subsequently reweights the importance of different channels:
\begin{equation}
\begin{split}
f_{mask}&=\mathrm{Sigmoid}( \mathrm{Conv} ( \mathrm{Concat} ([{f}_{\tau_{1}},\cdots,{f}_{\tau_{s}}])))\\
{f}_{out}&=f_{cat}+f_{cat} \times f_{mask}
\end{split}
\end{equation}
where $\tau_{s}$ represents the different kernel sizes used in the multi-scale strategy, and ${f}_{\tau_{s}}$ is the corresponding output of each dynamic convolution operation (see Fig. \ref{fig:overall_arch}). 

\begin{figure*}[h]
  \centering
    \includegraphics[width=0.8\linewidth]{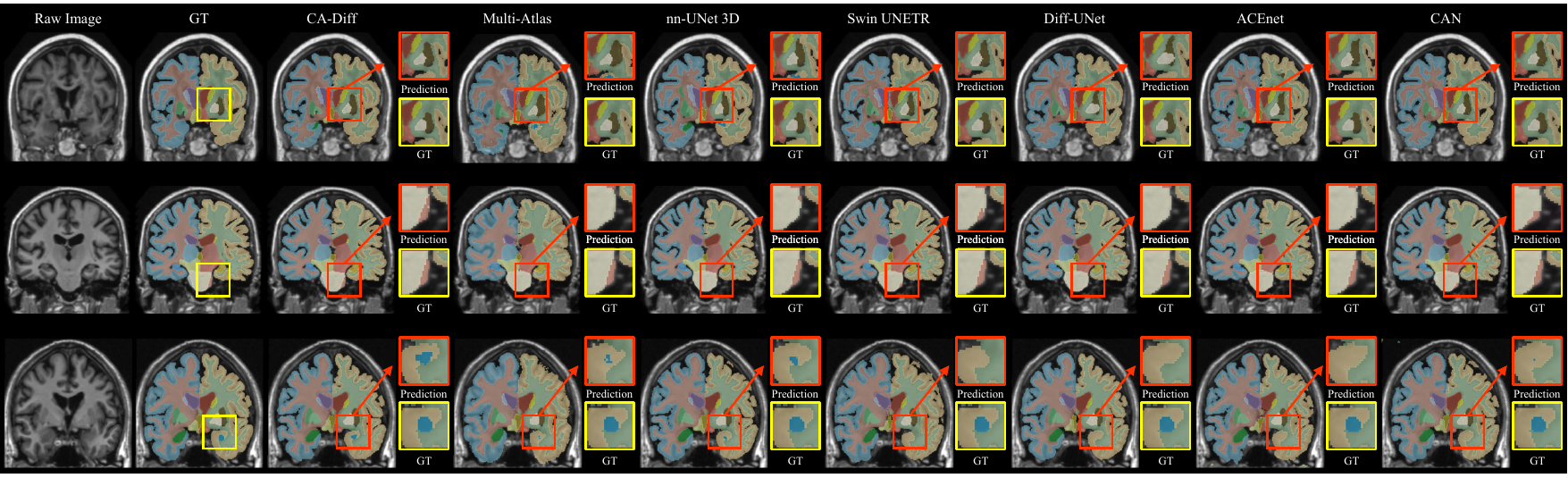}
    \caption{Qualitative visualizations of the proposed method and baseline models on the MALC dataset.}
    \label{fig:visual_compare}
  \hfill
\end{figure*}

\section{Experiments}
In this section, we conduct extensive experiments to demonstrate the effectiveness of our CA-Diff.

\begin{table*}[t]
\centering
    \begin{minipage}[c]{0.55\linewidth}
    \caption{Ablation of Framework Design.}\label{ab_model_design}
    \centering
    \scalebox{1}{
    \renewcommand\arraystretch{1.0}
                \begin{tabular}{lcccc}
                \hline
                \toprule[0.7pt]
                \multicolumn{1}{c}{\multirow{2}{*}{\textbf{Design}}} & \multicolumn{2}{c}{\textbf{MALC}} & \multicolumn{2}{c}{\textbf{SchizBull}} \\ \cline{2-5} 
                \multicolumn{1}{c}{}                         & Dice$\uparrow$        & NSD$\uparrow$         & Dice$\uparrow$          & NSD$\uparrow$           \\ \hline
                Diffusion                                    & 88.72       & 92.23       & 88.07         & 89.17         \\
                Collaborative Diffusion                           & 89.62       & 93.34       & 88.59         & 89.41         \\
                Collaborative Diffusion+$\mathcal{L}_{\mathrm{sac}}$          & 90.19       & 93.79       & 88.86         & 89.70         \\
                \rowcolor{gray!20}
                \textbf{Collaborative Diffusion+}$\bm{\mathcal{L}_{\mathrm{sac}}}$\textbf{+TACA}      & \textbf{90.63}       & \textbf{94.32}       & \textbf{89.29}         & \textbf{89.85}         \\ 
                \bottomrule[0.7pt]
                \hline
                \end{tabular}}

    \end{minipage}
    \hfill
    \begin{minipage}[c]{0.44\linewidth}
    \caption{Ablation of Fusion Method.}\label{ab_fuse}
    \centering
    \scalebox{1}{
    \renewcommand\arraystretch{1.0}
        \begin{tabular}{lcccc}
        \hline
        \toprule[0.7pt]
        \multirow{2}{*}{\textbf{Fusion Method}} & \multicolumn{2}{c}{\textbf{MALC}}        & \multicolumn{2}{c}{\textbf{SchizBull}}   \\ \cline{2-5} 
                                                & Dice$\uparrow$           & NSD$\uparrow$            & Dice$\uparrow$           & NSD$\uparrow$            \\ \hline
        Diffusion                               & 88.72          & 92.23          & 88.07          & 89.17          \\
        Dual condition                          & 89.03          & 92.61          & 88.27          & 89.25          \\
        \rowcolor{gray!20}
        \textbf{Collaborative Diffusion\footnotemark}             & \textbf{89.62} & \textbf{93.34} & \textbf{88.59} & \textbf{89.41} \\ 
        \bottomrule[0.7pt]
        \hline
        \end{tabular}}
    \end{minipage}
\end{table*}

\begin{table*}[t]
\centering
    \begin{minipage}[c]{0.33\linewidth}
    \caption{Ablation of Similarity Measure.}\label{ab_sim}
    \centering
    \scalebox{1}{
    \renewcommand\arraystretch{1.0}
                    \begin{tabular}{lcccc}
                    \hline
                    \toprule[0.7pt]
                    \multicolumn{1}{c}{\multirow{2}{*}{\textbf{Sim}}} & \multicolumn{2}{c}{\textbf{MALC}} & \multicolumn{2}{c}{\textbf{SchizBull}} \\ \cline{2-5} 
                    \multicolumn{1}{c}{}                                    & Dice$\uparrow$        & NSD$\uparrow$        & Dice$\uparrow$          & NSD$\uparrow$           \\ \hline
                    Dice                                                    & 90.40       & 94.17      & 89.08         & 89.70         \\
                    \rowcolor{gray!20}
                    \textbf{Cosine}                                                  & \textbf{90.63}       & \textbf{94.32}      & \textbf{89.29}         & \textbf{89.85}         \\ \bottomrule[0.7pt] \hline
                    \end{tabular}}
    \end{minipage}
    \begin{minipage}[c]{0.33\linewidth}
    \caption{Ablation of Condition.}\label{ab_cond}
    \centering
    \scalebox{1}{
    \renewcommand\arraystretch{1.0}
            \begin{tabular}{lcccc}
            \hline
            \toprule[0.7pt]
            \multicolumn{1}{c}{\multirow{2}{*}{\textbf{Cond}}} & \multicolumn{2}{c}{\textbf{MALC}} & \multicolumn{2}{c}{\textbf{SchizBull}} \\ \cline{2-5} 
            \multicolumn{1}{c}{}                          & Dice$\uparrow$        & NSD$\uparrow$        & Dice$\uparrow$          & NSD$\uparrow$           \\ \hline
            Label                          & 89.36       & 93.11      & 88.22         & 89.23         \\
            \rowcolor{gray!20}
            \textbf{$\mathbf{D}$\footnotemark[1]}                                            & \textbf{89.62}       & \textbf{93.34}      & \textbf{88.59}         & \textbf{89.41}         \\ 
            \bottomrule[0.7pt] \hline
            \end{tabular}}
    \end{minipage}
    \begin{minipage}[c]{0.33\linewidth}
    \caption{Ablation of Atlas.}\label{ab_atlas}
    \centering
    \scalebox{1}{
    \renewcommand\arraystretch{1.0}
            \begin{tabular}{lcccc}
            \hline
            \toprule[0.7pt]
            \multirow{2}{*}{\textbf{Atlas}} & \multicolumn{2}{c}{\textbf{MALC}} & \multicolumn{2}{c}{\textbf{SchizBull}} \\ \cline{2-5} 
                                   & Dice$\uparrow$        & NSD$\uparrow$        & Dice$\uparrow$          & NSD$\uparrow$           \\ \hline
            
            Sample                 & 90.26       & 94.09      & 89.05         & 89.68         \\
            MNI                 & 90.42       & 94.13      & 89.18         & 89.72         \\
            \rowcolor{gray!20}
            \textbf{Colin27}                & \textbf{90.63}       & \textbf{94.32}      & \textbf{89.29}         & \textbf{89.85}         \\
            \bottomrule[0.7pt] \hline
            \end{tabular}}
    \end{minipage}
\end{table*}

\subsection{Datasets}
We conduct experiments on the following datasets.

\textbf{MALC Dataset.}
The 2012 Multi-Atlas Labeling Challenge (MALC) features MRI T1 brain scans from 30 individuals, with the entire brain manually segmented into 27 coarse-grained structures\cite{MALC}. 

\textbf{SchizBull Dataset.}
The SchizBull 2008 dataset contains 103 subjects. 
The dataset is manually annotated with 32 brain structures and is part of the CANDI dataset~\cite{CANDIShare}.

\textbf{Hammers Dataset.}
The Hammers brain dataset \cite{hammers_49} contains 30 MRI scans.
The segmentation map includes 95 brain structures.

\subsection{Implementation Details}

We construct the distance field using the Colin27 brain atlas \cite{colin} as the reference image, with registration performed via the ANTs package.
For 3D tasks, we use a patch size of $96 \times 96 \times 96$ and a batch size of 2, while 2D tasks adopt $160 \times 160$ patches with a batch size of 70. The TACA module employs kernel sizes of 7, 11, and 15. The model is trained for 200,000 iterations using the AdamW optimizer with a learning rate of $1e^{-4}$. 
Inference uses five diffusion steps, and three-fold cross-validation is performed. Evaluation metrics include Dice and NSD, with all methods rerun for fairness without post-process. 

\subsection{Comparison with SOTAs}

Table \ref{result_compare} presents the average Dice coefficient and NSD results, including standard deviations. On the MALC dataset, CA-Diff surpasses the 3D nnU-Net baseline by 2.01\% in Dice and 1.67\% in NSD. Additionally, it outperforms the state-of-the-art CAN model, which incorporates skull boundary information as a prior.  
CA-Diff achieves a 0.91\% improvement in Dice on the SchizBull dataset and a 1.12\% improvement on the Hammers dataset compared to the CAN model, with superior surface overlap in both cases. Segmentation results on the MALC dataset are visualized in Fig.~\ref{fig:visual_compare}.

\subsection{Ablation Study}

\textbf{Framework Design.}
CA-Diff incorporates several designs to enhance segmentation (Table \ref{ab_model_design}). The collaborative diffusion method leverages the distance field to guide the model, improving accuracy. Consistency loss further refines spatial-anatomical relationships, while the TACA module adaptively reweights channels in skip connections, providing additional performance gains.

\textbf{Fusion Method.}
We perform a comparative analysis with the \textit{Dual Condition} method. 
The results summarized in Table \ref{ab_fuse} highlight the superior performance of our proposed method.

\textbf{Similarity Measure.}
We compare different choice of similarity measurement when calculating the consistency loss.
As shown in Table \ref{ab_sim}, we compare the cosine similarity and the Dice score, 
and cosine similarity proves to be more effective. 

\footnotetext{The optimal values differ from those in Table \ref{result_compare} because the model design is incompatible with the ablation setting. As a result, the consistency loss and TACA module are excluded to ensure fair comparison.}

\textbf{Condition Type.}
To compare with label-like conditions that record anatomical classes, we randomly select sample labels and transform them using the registration method to serve as the conditions for other samples. This procedure is repeated three times for each cross-validation fold.
As shown in Table \ref{ab_cond}, the model using the distance field shows better accuracy, indicating the usefulness of fusing global features.

\textbf{Atlas.}
We evaluate the sensitivity of the atlas used in distance field construction (Table \ref{ab_atlas}). In addition to the Colin27 atlas, we use random samples (the procedure is repeated three times) and the MNI atlas \cite{mazziotta1995probabilistic}. 
Results indicate that the high-resolution Colin27 atlas provides better performance.

\begin{figure}[t]
  \centering
   \includegraphics[width=0.8\linewidth]{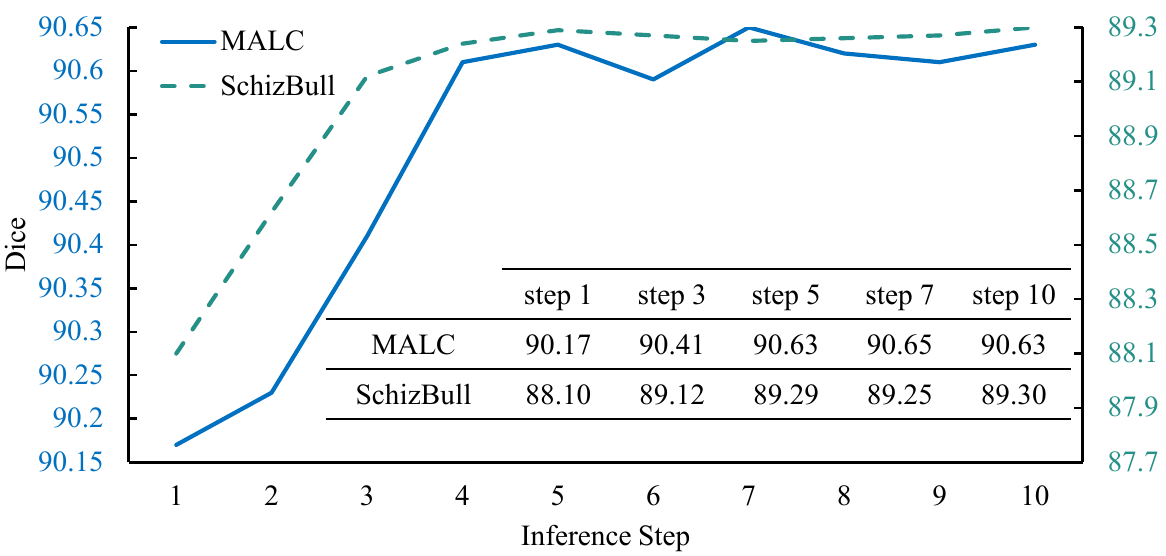}
   \caption{Ablation of inference step.}
   \label{fig:StepEffect}
\end{figure}

\textbf{Inference Step.}
The effect of varying the number of inference steps on the Dice score is illustrated in Fig.~\ref{fig:StepEffect}, revealing that the CA-Diff progressively enhances performance with an increase in steps and reaches convergence within five steps. 

\section{Conclusion}
We propose CA-Diff for brain tissue segmentation from MRI scans, introducing a distance field as an anatomical condition. To leverage this, we develop a collaborative denoising method to link spatial locations with anatomical structures, enhanced by a consistency loss and a time-adapted channel attention module. 
While CA-Diff demonstrates significant potential, its performance may be influenced by the precision of the registration method used. 
Additionally, the iterative inference procedure introduces some computational overhead. 
Nevertheless, integrating specific anatomical features as conditional inputs through the parallel denoise method in diffusion models tailored for diverse anatomical structures emerges as a promising avenue for future exploration.

\section*{Acknowledgment}
This work is supported by the National Natural Science Foundation of China (NSFC No. 62272184 and No. 62402189), the China Postdoctoral Science Foundation under Grant Number GZC20230894, the China Postdoctoral Science Foundation (Certificate Number: 2024M751012), and the Postdoctor Project of Hubei Province under Grant Number 2024HBBHCXB014. The computation is completed in the HPC Platform of Huazhong University of Science and Technology.

\bibliographystyle{IEEEbib}
\bibliography{references}

\vspace{12pt}

\end{document}